\documentclass[12pt]{article}
\usepackage{amssymb,amsmath}
\usepackage[noblocks]{authblk}
\usepackage[top=0.75in, bottom=0.75in, left=0.75in, right=0.75in, dvips]{geometry}
\usepackage{caption}
\begin{document}
\textwidth 10.0in 
\textheight 9.0in 
\topmargin -0.60in
\title{Renormalization Group Summation with Heavy Fields}
\author[1,2]{D.G.C. McKeon}
\affil[1] {Department of Applied Mathematics, The
University of Western Ontario, London, ON N6A 5B7, Canada} 
\affil[2] {Department of Mathematics and
Computer Science, Algoma University, \newline Sault St.Marie, ON P6A
2G4, Canada}
\date{}
\maketitle        

\maketitle
\noindent
email: dgmckeo2@uwo.ca

\begin{abstract}
The summation of logarithmic contributions to perturbative radiative corrections in physical processes through use of the renormalization group equation has proved to be a useful way of enhancing the information one can obtain from explicit calculation.  However, it has proved difficult to perform this summation when massive fields are present.  In this note we point out that if the masses involved are quite large, the decoupling theorem of Symanzik and of Appelquist and Carazzone can be used to make the summation of logarithms possible.
\end{abstract}

\section{Introduction}

Higher loop calculations in perturbative quantum chromodynamics (QCD) lead to results that depend on the logarithm of the unphysical renormalization mass scale $\mu^2$, typically of the form $\ln\left(\frac{s}{\mu^2}\right)$ where $s$ is the centre of mass energy in the process being considered.  As $\mu^2$ is unphysical, one has the renormalization group (RG) equation [1-5] which in many instances makes it possible to sum logarithmic corrections [6,7].  However, when massive fields are present, the form of the logarithms that arise is often so complicated that such summation is not feasible.

In the instance that the masses $M^2$ that arise are much greater than the energy scale of the process, a decoupling theorem due to Symanzik [8] and Appelquist and Carazzone [9] states that up to order $1/M^2$, these masses serve only to renormalize the parameters that characterize the full theory, leading to an effective low energy theory in which these massive fields are not present (ie, they ``decouple'').  As both the full theory and the effective low energy theory satisfy the RG equation, it proves possible to relate the running parameters of the full theory with those of the effective theory when one employs a mass independent renormalization scheme [10,11].

This result has the consequence of making it possible to perform RG summation of logarithmic contributions  to radiative effects in the massless effective theory, and then to incorporate the contribution of heavy fields by invoking the Appelquist-Carazzone-Symanzik (ACS) theorem.  In ref. [15] the technique of RG summation was applied to the inclusive semileptonic decays of $b \rightarrow u$.  In this case, the mass of the $b$ quark appears explicitly and all other quarks ($u, d, s, c$) are taken to be massless.  In this paper we will consider the process $(e^+e^- \rightarrow$ hadrons) and show how a single massive quark can implicitly affect the strong couplant $a$ using the ACS theorem.

\section{Summation of Logarithms and the ACS theorem}

Let us denote the amplitude for the process $e^+e^- \rightarrow \rm{hadrons}$ in the full theory by $\Gamma(s, a, m, \mu)$ and in the effective low energy theory (in which $s \ll m^2$) by $\Gamma^*(s, a^*, \mu)$.  The ACS theorem implies that [10]
\begin{equation}
\Gamma^*(s,a^*, \mu) = Z \Gamma(s, a, m, \mu) + O \left(\frac{1}{m^2}\right)
\end{equation}
where $a$ and $a^*$ are the couplings in the two theories, and
\begin{subequations}
\begin{align}
Z  &= Z(a,m/\mu) \\
a^* &= a^*(a,m/\mu).
\end{align}
\end{subequations}
Since $\mu$ is an unphysical parameter, we have the RG equations
\begin{subequations}
\begin{align}
\mu \frac{d\Gamma}{d\mu} &= \left( \mu \frac{\partial}{\partial \mu} + \beta (a) \frac{\partial}{\partial a} + m \delta (a) \frac{\partial}{\partial m} \right) \Gamma\\
& = 0 \nonumber \\
&= \frac{d\Gamma^*}{d\mu} = \left( \mu \frac{\partial}{\partial \mu} + \beta^* (a^*) \frac{\partial}{\partial a^*} \right) \Gamma^* \; ,
\end{align}
\end{subequations}
where
\begin{subequations}
\begin{align}
\beta (a) &= \mu \frac{da}{d\mu} = -b a^2 (1 + ca + c_2a^2 + \ldots ) \\
\beta^* (a^*) &= \mu \frac{da^*}{d\mu} = -b^* a^{*^{2}} (1 + c^*a^* + c_2^*a^{*^{2}} + \ldots )
\end{align}
\end{subequations}
and
\begin{equation}
m\delta  (a) = \mu \frac{dm}{d\mu} = mfa(1 + g_1a + g_2a^2 + \ldots ) .
\end{equation}
We are using the mass independent renormalization scheme [4,5] with minimal subtraction [4].

Using the conventions of refs. [13, 15], we will characterize the boundary conditions to eqs. (4, 5) by massive parameters $\Lambda$, $\Lambda^*$ and $I\!\!M$, writing
\begin{subequations}
\begin{align}
\ln \left(\frac{\mu}{\Lambda}\right) &= \int_0^{a(\mu)} \frac{dx}{\beta(x)} + \int_0^\infty \frac{dx}{bx^2(1+cx)}\\
\ln \left(\frac{\mu}{\Lambda^*}\right) &= \int_0^{a^*(\mu)} \frac{dx}{\beta(x)} + \int_0^\infty \frac{dx}{b^*x^2(1+c^*x)}\\
\end{align}
\end{subequations}
and
\begin{equation}
m(\mu) = I\!\!M \exp \left[\int_0^{a(\mu)} \frac{dx\delta(x)}{\beta(x)} + \int_0^\infty \frac{dx fx}{bx^2(1+cx)}\right].
\end{equation}

Together, eqs. (1-3) lead to [10]
\begin{align}
\beta^* (a^*) &= \left( \mu \frac{\partial}{\partial\mu} + \beta (a) \frac{\partial}{\partial a} + m \delta (a) \frac{\partial}{\partial m} \right) a^* (a, \mu/m )\\
&= \left( (1 - \delta(a) \frac{\partial}{\partial t} + \beta (a) \frac{\partial}{\partial a}\right) a^*(a,t); \qquad   \left( t \equiv \ln \frac{\mu}{ m} \right)\nonumber
\end{align}
this equation relates the couplings $a$ and $a^*$ in the low energy region.  

If now the expansion
\begin{equation}
\Gamma^* = \sum_{n=0}^\infty \sum_{m=0}^n T_{nm}^* L^m a^{*n+1}
\end{equation}
$(L \equiv \ln (\mu / \sqrt{s}))$ is substituted into eq. (3b), then
\begin{equation}
A_n (a^*) = - \frac{\beta^*(a^*)}{n} \frac{d}{da^*} A_{n-1} (a^*)
\end{equation}
where
\begin{equation}
A_n (a^*) = \sum_{m=0}^\infty T_{n+m,n}^* a^{*n+m+1} .
\end{equation}
Using eq. (4b), eq. (10) leads to
\begin{align}
A_n(a^*(\eta)) &= -\frac{1}{n} \mu\frac{d}{d\mu} A_{n-1} (a^*(\mu)) \nonumber \\
&= \frac{(-1)^n}{n!} \left(\mu \frac{d}{d\mu}\right)^n A_0 (a^*(\mu))
\end{align}
and so by eqs. (9-12),
\begin{align}
\Gamma^* (s, a^*, \mu) &= \sum_{n=0}^\infty \frac{(-L)^n}{n!}\left(\mu\frac{d}{d\mu}\right)^n A_0(a^*(\mu))\nonumber \\
&= A_0 \left( a^*\left( \ln \frac{\mu}{\Lambda^*} - L\right)\right)\nonumber\\
& = A_0 \left( a^*\left( \ln \frac{\sqrt{s}}{\Lambda^*}\right)\right).\nonumber
\end{align}

By eqs. (11, 12) we find that all dependence of $\Gamma^*$ on $\mu$ has cancelled, leaving
\begin{equation}
\Gamma^* = \sum_{n=0}^\infty T_n^* A_0 \left( a^*\left( \ln \left( \frac{\sqrt{s}}{\Lambda^*} \right)\right)\right)^{n+1} \qquad (T_n^* \equiv T_{n0}^*).
\end{equation}
In ref. [13] it is shown that renormalization scheme ambiguities when using mass independent renormalization can be characterized by $c_2^*, c_3^* \ldots$; in ref. [12] it is shown how the requirement that $\frac{d\Gamma^*}{dc_i^*} = 0$ results in $T_n^*$ being expressed as 
\begin{subequations}
\begin{align}
T_0^* &= 1\\
T_1^* &= \tau_1^* \\
T_2^* &= -c_2^* + \tau_2^* \\
T_3^* &= -2c_2^* \tau_1^* - \frac{1}{2} c_3^* + \tau_3^*
\end{align}
\end{subequations}
etc. where $\tau_i^*$ are all renormalization scheme invariants.

We now return to eq. (8) which relates the running coupling $a^*$ in the effective low energy theory to the running coupling $a$ in the full theory. This can be accomplished by employing the ``method of characteristics'' outlined in ref. [16].  The first step is to define auxiliary functions $\overline{a}(\sigma)$, $\overline{a}^*(\sigma)$, and $\overline{t}(\sigma)$ such that
\begin{subequations}
\begin{align}
\frac{d\overline{a}(\sigma)}{d\sigma} & = \beta (\overline{a}(\sigma))\\
\frac{d\overline{a}^*(\sigma)}{d\sigma} & = \beta^* (\overline{a}(\sigma))\\
\intertext{and}
\frac{d\overline{t}(\sigma)}{d\sigma} & = 1-\delta (a(\sigma)).
\end{align}
\end{subequations}
It follows from eq. (15) that
\begin{equation}
\left[\left( 1 - \delta(\overline{a}(\sigma))\right) \frac{\partial}{\partial\overline{t}(\sigma)} + \beta(\overline{a}(\sigma)) 
\frac{\partial}{\partial\overline{a}(\sigma)}\right] \overline{a}^*\left( \overline{a}(\sigma , \overline{t}(\sigma)\right) = \beta^*(\overline{a}^*(\sigma)).
\end{equation}
We can write the solutions to eq. (15) as
\begin{subequations}
\begin{align}
\sigma &= \int_0^{\overline{a}(\sigma)} \frac{dx}{\beta(x)} + \int_0^\infty \frac{dx}{bx^2(1 + cx)} + c_1(\tau)\\
\sigma &= \int_0^{\overline{a}^*(\sigma)} \frac{dx}{\beta^*(x)} + \int_0^\infty \frac{dx}{b^*x^2(1 + c^*x)} + c_2(\tau)\\
\overline{t}(\sigma) &= \int_0^{\overline{a}(\sigma)} dx \frac{1-\delta(x)}{\beta(x)} + \int_0^\infty dx \frac{1-fx}{bx^2(1 + cx)} + c_3(\tau)
\end{align}
\end{subequations}
where $c_1(\tau)$, $c_2(\tau)$, $c_3(\tau)$ are a set of boundary conditions characterized by a parameter $\tau$.  By eqs. (17a,b) we find that
\begin{equation}
\int_0^{\overline{a}^*(\sigma) }\frac{dx}{\beta^*(x)} + \int_0^\infty \frac{dx}{b^*x^2(1 + c^*x)} = \int_0^{\overline{a}(\sigma)} \frac{dx}{\beta(x)} + \int_0^\infty \frac{dx}{bx^2(1 + cx)}+ c_2(\tau) - c_1(\tau) .
\end{equation}
If $\phi(x) = c_2\left(c_3^{-1}(x)\right) - c_1\left(c_3^{-1}(x)\right)$ then together eqs. (17c, 18) result in the general solution to eq. (16) written so that $\overline{a}^*(\sigma)$ implicitly depends on $\sigma$ through $a(\sigma)$ and $t(\sigma)$,
\begin{align}
\int_0^{\overline{a}^*(\sigma)} \frac{dx}{\beta^*(x)}+ \int_0^\infty  \frac{dx}{b^*x^2(1 + c^*x)} &= \int_0^{\overline{a}(\sigma)} \frac{dx}{\beta(x)} + \int_0^\infty \frac{dx}{bx^2(1 + c^*x)}  \\
&+ \phi\left( \overline{t}(\sigma) - \int_0^{\overline{a}(\sigma)} dx
\frac{1-\delta(x)}{\beta(x)} -\int_0^\infty dx \frac{1-fx}{bx^2(1 + cx)} \right).\nonumber
\end{align}
(Explicit differentiation of eq. (19) shows that $\overline{a}^*(\overline{a}(\sigma), \overline{t}(\sigma))$ satisfies eq. (16) for arbitrary $\phi$.)

If we now impose the boundary conditions
\begin{equation}\tag{20a-c}
\overline{a}(0) = a, \qquad  \overline{a}^*(0) = a^*, \qquad \overline{t}(0) = t
\end{equation}
on eq. (15), then setting $\sigma = 0$ in eq. (19) and then using eqs. (6, 7) in eq. (19) leads to 
\begin{equation}\tag{21}
\ln \left( \frac{\mu}{\Lambda^*}\right) = \ln \left( \frac{\mu}{\Lambda}\right) + \phi 
\left( \ln  \frac{\mu}{I\!\!M} - \ln \left(\frac{\mu}{\Lambda}\right)\right)
\end{equation}
or
\begin{equation}\tag{22a}
\phi \left( \frac{\Lambda}{I\!\!M} \right) = \ln \left( \frac{\Lambda}{\Lambda^*} \right) = \ln \left( \frac{\Lambda}{I\!\!M} \right) + \ln \left( \frac{I\!\!M}{\Lambda^*} \right) .
\end{equation}
As $ \ln \left( \frac{\Lambda}{I\!\!M} \right) $ is not some fixed number, we conclude that 
\begin{equation}\tag{22b}
\phi(x) = x + \ln\left( \frac{I\!\!M}{\Lambda^*} \right) .
\end{equation}
We thus see that the boundary conditions of eq. (20) fixes the function $\phi$ in eq. (19) so that
\begin{equation}\tag{23}
\int_0^{a^*} \frac{dx}{\beta^*(x)} + \int_0^\infty \frac{dx}{b^*x^2(1+cx)} = \
t + \int_0^a \frac{dx\delta(x)}{\beta(x)} + \int_0^\infty dx \frac{fx}{bx^2(1+cx)} + \ln\left(\frac{I\!\!M}{\Lambda^*}\right).
\end{equation}
In eq. (23) we have found an implicit form for the function $a^*(t,a)$ that is the most general solution to eq. (8) which is consistent with the boundary conditions of eqs. (6, 7).  This approach to using the method of characteristics is different to what was employed in ref. [17].

The question remains of the renormalization schemes used to compute $\Gamma^*$ and $\Gamma$ respectively.  (This has been considered in the context of supersymmetry in ref. [14].)  In principle these two choices can be made independently; there is no relationship between $a$ and $a^*$ that fixes how the parameters $c_i$ of eq. (4a) and $g_i$ of eq. (5) are related to the parameters $c_i^*$ of eq. (4b).  The values of $c_i^*$ affect $T_n^*$ through eq. (14); however, altering the values of $c_i$ and $g_i$ only serves to alter the relationship of $a$ and $m$ with $\mu$. But as $\Gamma^*$ is independent of $\mu$, it is possible to select whatever value of $c_i$ and $g_i$ that is most convenient.  For example, if $c_i = g_i = 0$, then eq. (23) is greatly simplified.  The choice $c_i^* = 0$ further simplifies eq. (23); it also reduces $a^*$ in eq. (6b) to being a Lambert $W$ function [12].  One could also choose $c_i^*$ so that $T_n^* = 0\quad (n \geq 2)$, reducing the sum in eq. (13) to being just two terms.

If we work in the `t Hooft scheme, both when computing $\Gamma$ and $\Gamma^*$, then $c_i = g_i = c_i^* = 0$ and eq. (23) leads to 
\begin{equation}\tag{24}
\left( 1 + \frac{1}{c^*a^*}\right)\exp \left( 1 + \frac{1}{c^*a^*}\right) = 
\left( 1 + \frac{1}{ca}\right)^{-fb^*/bc^*} 
\left( \frac{I\!\!M}{\Lambda^*}\right)^{-b^*/c^*} \exp \left( 1 -\frac{b^*t}{c^*}\right)
\end{equation}
from which we can obtain $a^*(a,t)$ using the Lambert function $W(x)$ that is defined by $W(x)\exp W(x) = x$ [18].

In ref. [15] it is shown that the parameter $I\!\!M$ introduced in eq. (7) is a renormalization scheme invariant.  Furthermore, it is shown there that $I\!\!M$ is related to the pole (``physical'') mass $m_{\rm{pole}}$ by the equation
\begin{equation}\tag{25}
m_{\rm{pole}} = I\!\!M \exp \left[ \int_0^{a\left(\ln\frac{m_{\rm{pole}}}{\Lambda}\right)} dx \frac{\delta(x)}{\beta(x)} + \int_0^\infty dx 
\frac{fx}{bx^2(1+cx)}\right] \sum_{k=0}^\infty  \kappa_{k,0} a^k \left(\ln \frac{m_{\rm{pole}}}{\Lambda}\right).
\end{equation}
The constants $\kappa_0$, $\kappa_1$, $\kappa_2$ have been computed in ref. [19] by evaluating the massive Fermion self energy to two loop order.  Eq. (25) can be used to eliminate $I\!\!M$ in eq. (24) in favour of $m_{\rm{pole}}$.

\section{Conclusion}

We have combined RG summation with the ACS theorem to show how the calculation of the amplitude for the process $e^+e^- \rightarrow \rm{hadrons}$ can take into account the presence of heavy field.  We have also demonstrated how the result is independent of the renormalization scale $\mu$ and how one can make convenient choices for the parameters $c_i^*$, $c_i$ and $g_i$ that parameterize the renormalization schemes chosen.

\section*{Acknowledgements}
R. Macleod and D. Wheatley assisted in this work.


\begin{thebibliography}{99}
\bibitem{1} E.C.G. Stueckelberg and A. Peterman, \textit{Helv. Phys. Acta} \textbf{26}, 499 (1953).
\bibitem{2} N.N. Bogolivbov and D.V. Shirkov, ``Introduction to the Theory of Quantized and Fields'' (Interscience, New York, 1959).
\bibitem{3} M. Gell-Mann and F. Low, \textit{Phys. Rev.} \textbf{95}, 1300 (1954).
\bibitem{4}  G. 't Hooft,  \textit{Nucl. Phys.} \textbf{B61}, 455 (1973).
\bibitem{5} S. Weinberg, \textit{Phys. Rev.} \textbf{D8}, 3497 (1973).
\bibitem{6} M.R. Ahmady, F.A. Chishtie, V. Elias, A.H. Fariborz, N. Fattahi, D.G.C. McKeon, T.N. Sherry and T.G. Steele, \textit{Phys. Rev.} \textbf{D66}, 014010 (2002).
\bibitem{7} M.R. Ahmady, F.A. Chishtie, V. Elias, A.H. Fariborz, D.G.C. McKeon, A. Squires and T.G. Steele, \textit{Phys. Rev.} \textbf{D67}, 034017 (2003).
\bibitem{8}  K. Symanzik,  \textit{Comm. Math. Phys.} \textbf{34}, 7 (1973).
\bibitem{9}  T. Appelquist and J. Carazzone,  \textit{Phys. Rev.} \textbf{D11}, 2856 (1975).
\bibitem{10} E. Witten,  \textit{Nucl. Phys.} \textbf{B104}, 445 (1976).
\bibitem{11} E. Witten,  \textit{Nucl. Phys.} \textbf{B122}, 109 (1977).
\bibitem{12} D.G.C. McKeon, \textit{Phys. Rev.} \textbf{D92}, 045031 (2015).
\bibitem{13} P.M. Stevenson, \textit{Phys. Rev.} \textbf{D23}, 2916 (1981).
\bibitem{14} V. Elias and D.G.C. McKeon, \textit{Can. J. Phys.} \textbf{84}, 131 (2006).
\bibitem{15} F.A. Chishtie, D.G.C. McKeon and T.N. Sherry, arxiv hep-ph 1708.04219.
\bibitem{16} R. Courant and D. Hilbert, ``Methods of Mathematical Physics, Vol. II, Chapter II'' (John Wiley and Sons, N.Y., 1962).
\bibitem{17} W. Weisberger, \textit{Phys. Rev.} \textbf{D24}, 1617 (1981).
\bibitem{18} E. Gardi, G. Grunberg and M. Karliner, \textit{JHEP} \textbf{9807:007},  (1998)\\
R.M. Corless, G.H. Gonnet, D.E.G. Hare, D.J. Jeffrey and D.E. Knuth,  \textit{Adv. Comput. Math.} \textbf{5}, 329 (1996).
\bibitem{19} N.G. Grey, D.J. Broadhurst, W. Grafe and K. Schilcher,  \textit{Z. Phys. } \textbf{C48}, 673 (1990)\\
D.J. Broadhurst, N. Grey and K. Schilcher,  \textit{Z. Phys.} \textbf{C52}, 111 (1991)\\
J. Feischer, F. Jegerlehner, O.V. Tarasov and O.L. Vevetin,  \textit{Nucl. Phys.} \textbf{B539}, 671 (1999)\\
K.G. Chatyrkin and M. Steinhauser,  \textit{Nucl. Phys.} \textbf{B573}, 617 (2000)\\
K. Melnikov and T. van Ritbergen,  \textit{Phys. Lett.} \textbf{B482}, 99 (2000)\\
L.V. Avdeev and M.Y. Kalmykov,  \textit{Nucl. Phys.} \textbf{B502}, 419 (1997)\\
A.L. Kataev and V.S. Molokoedov, \textit{Eur. J. Phys.} \textbf{131}, 271 (2016).


\end{thebibliography}
\end{document}